\documentclass[english,prl,twocolumn,floatfix]{revtex4}
\usepackage{graphicx}  
\usepackage{dcolumn}   
\usepackage{bm}        
\usepackage{amssymb}   
\usepackage{amsmath}

\usepackage[T1]{fontenc}
\usepackage[latin9]{inputenc}
\usepackage{amsthm}
\usepackage{amsmath}
\usepackage{graphicx}
\usepackage{color}

\makeatletter
\@ifundefined{textcolor}{}
{%
 \definecolor{BLACK}{gray}{0}
 \definecolor{WHITE}{gray}{1}
 \definecolor{RED}{rgb}{1,0,0}
 \definecolor{GREEN}{rgb}{0,1,0}
 \definecolor{BLUE}{rgb}{0,0,1}
 \definecolor{CYAN}{cmyk}{1,0,0,0}
 \definecolor{MAGENTA}{cmyk}{0,1,0,0}
 \definecolor{YELLOW}{cmyk}{0,0,1,0}
 }

\makeatletter

\newcommand{\Rmnum}[1]{\expandafter\@slowromancap\romannumeral #1@}
\makeatother

\newcommand{\be}{\begin{equation}}
\newcommand{\ee}{\end{equation}}

\def\lsim{\mathrel{\rlap{\lower4pt\hbox{$\sim$}}
    \raise1pt\hbox{$<$}}}                
\def\gsim{\mathrel{\rlap{\lower4pt\hbox{$\sim$}}
    \raise1pt\hbox{$>$}}}

\renewcommand\[{\begin{equation}}
\renewcommand\]{\end{equation}}
\usepackage[english]{babel}
\addto\captionsenglish{}
\makeatother

\usepackage{babel}

\begin{document}
\bibliographystyle{unsrt}
\title{Phenomenlogical Modeling of Durotaxis}

\author{Guangyuan Yu$^{1,2}$, Jingchen Feng$^{2}$, Haoran Man$^{1}$,Herbert Levine$^{2,3}$}

\affiliation{
$^{1}$ Physics and Astronomy Department, Rice University, Houston, USA\\
$^{2}$ Center for Theoretical Biological Physics, Rice University, Houston, USA\\
$^{3}$ Bioengineering Department, Rice University, Houston, USA }

\date{\today}
\begin{abstract}
Cells exhibit qualitatively different behaviors on substrates with different rigidities. The fact that cells are more polarized on the stiffer substrate motivates us to construct a two-dimensional cell with the distribution of focal adhesions dependent on substrate rigidities. This distribution affects the forces exerted by the cell and thereby determines its motion. Our model reproduces the experimental observation that the persistence time is higher on the stiffer substrate. This stiffness dependent persistence will lead to durotaxis, the preference in moving towards stiffer substrates. This propensity is characterized by the durotaxis index first defined in experiments. We derive and validate a 2D corresponding Fokker-Planck equation associated with our model. Our approach highlights the possible role of the focal adhesion arrangement in durotaxis.
\end{abstract}
\maketitle 
Cells are capable of sensing and responding to the mechanical properties of their external environment. For example, cytoskeletal stiffness \cite{WangButlerIngberEtAl1993a}, cellular differentiation \cite{GuilakCohenEstesEtAl2009,ParkChuTsouEtAl2011,EvansMinelliGentlemanEtAl2009,TrappmannGautrotConnellyEtAl2012} and cell morphology and motility \cite{PelhamWang1997,YeungGeorgesFlanaganEtAl2005,LoWangDemboEtAl2000} are all strongly influenced by ECM stiffness. In particular, it has been shown experimentally that cells prefer crawling towards the stiffer parts on substrates with spatially varying rigidity, a property which is referred to as durotaxis.  Durotaxis is a universal property of motile cells, despite the diverse shapes and structures among different cell types. It has been proposed that durotaxis is critical for fine-tuning cell path-finding and wound healing \cite{Basan2013,Losick2013}. Also, there is increasing evidence showing that durotaxis is involved in cancer metastasis,  since tumors are usually stiffer than the surrounding materials \cite{Huang2005,Venkatesh2008}.

A standard approach to modeling cell motility is to assume that cells execute a persistent random walk \cite{SambethBaumgaertner2001,HuangBrangwynneParkerEtAl2005,SadjadiShaebaniRiegerEtAl2015a}; sometimes  Lévy walks are used instead \cite{Reynolds2010,BuldyrevGoldbergerHavlinEtAl1993}. Recently, Elizavata and colleagues applied persistent random walk ideas to understand durotaxis by relating persistence to substrate stiffness \cite{NovikovaRaabDischerEtAl2015}. Their approach did show how this assumption could lead to durotaxis, but did not propose any direct mechanical reason for this correspondence; also they did not fully analyze their model in the relevant case of a two-dimensional spatial domain.  In this study, we propose a simple intracellular mechanism that naturally leads to stiffness dependent persistence which, in agreement with the above findings, results in durotaxis. Our approach combines direct simulations with the derivation of a quantitatively accurate 2D Fokker-Planck equation, for which the numerical solution matches well with simulation data.

Our basic hypothesis is built on the fact that cells are observed to be more polarized when they move on stiffer materials. Cells have sophisticated mechanisms to sense stiffness, involving various cellular components and subsystems including the plasma membrane \cite{FacklerGrosse2008,Keren2011}, actin filaments \cite{GalkinOrlovaEgelman2012,MattilaLappalainen2008}, actomyosin-based contractility, integrin-based focal adhesions  \cite{KimWirtz2013,PlotnikovPasaperaSabassEtAl2012}, etc.  Once cells sense a stiffer substrate, they take on a more elongated shape \cite{trichet2012evidence,IsenbergDiMillaWalkerEtAl2009} as a response. Now,  cells move by protrusion which occurs with the help of focal adhesions which allow force transmission to the substrate. We will assume that the change in shape to being more polarized implies  that focal adhesions (FA) are formed within a narrower wedge on the cell front. It is also possible that the total number of FAs present at some fixed time increases on stiffer substrates, as FA are observed to be more stable on stiffer substrates \cite{PelhamWang1997}. In our model, both the distribution and the total number of FA directly control the variance of deflection angles in cell motion over a short time interval. This mechanism will create the necessary relationship between stiffness and persistence.

In experiments, the locations of cells moving on a 2D surface are typically recorded at fixed time intervals. Accordingly, we model the cell as a rigid object moving with velocity $v$ and rotating its motion direction $\Phi$ (its polarization) by an angle $\Delta\Phi$ at fixed time intervals $\Delta t=t_{i+1}-t_i$ which we take to be our unit of time. To determine $\Delta\Phi$, we assume there are a number $N_f$ of  focal adhesions which are positioned at distances $r_i$ from the cell center and angles $\theta_i$   relative to the current direction of motion; these are chosen randomly from uniform distributions with ranges $(r_{min},r_{max})$, $(-\theta_{max}, \theta_{max})$ respectively.
We assume in line with the previous arguments that $\theta_{max}$ is determined by local substrate stiffness $k$ as $\theta_{max} = A/k$, where $A$ is a constant factor. The basic picture of our cell is given in Fig. 1, Our calculations will assume that $N_f$ remains constant. The driving force from each focal adhesion is assumed to have a constant magnitude and to point in the current moving direction. The net driving force is canceled by the friction acting on the cell, thereby determining the velocity. At each time step, the dynamical formation and disruption of  FAs cause a possible imbalance in the driving torque. With fast relaxation, the cell will rotate by an angle $\Delta\Phi$ at each time step to satisfy zero net torque:
\be
\sum_{i=1}^{N_f}r_i \sin (\theta_i-\Delta\Phi)=0
\ee
whose solution is:
\be
\tan{\Delta\Phi}=\frac{\sum_{i=1}^{N_f}r_i \sin (\theta_i)}{\sum_{i=1}^{N_f}r_i \cos (\theta_i)}
\ee
For the purpose of illustration, we typically set $r_i=1$ for all $i$ and $N_f=12$ in our model. 

Clearly the variance of the induced distribution for $\Delta \Phi$ determines the persistence of the motion. Here
we use a Monte Carlo sampling method to evaluate this variance. We use $10^6$ sampling steps and have checked that this gives us an accurate evaluation for the range of parameters we have investigated. For the case of fixed radii, we obtain
\begin{equation}\label{eq4}
\int \cdots \int _{-\theta_{max}}^{\theta_{max}} \left( \prod_{i=1}^{N_f} \frac{d\theta_i}{2\theta_{max}} \right) \arctan^{2} \left( \frac{\sum_{i=1}^{N_f}\sin (\theta_{i})}{\sum_{i=1}^{N_f}\cos (\theta_{i})} \right)
\nonumber
\end{equation}
One can also compute the variance for the more general situation with a distribution for the radii as well. Typical results of this calculation are shown in Fig 2a. For use later on, we have fitted the data for the case $N_f=12$ with fixed radii to a simple function of $k$. 

\be
\sigma(k)=\frac{1}{\alpha k + \beta}
\ee

As expected, increasing $N_f$ or decreasing $\theta_{max}$ reduces the variance. Thus, rigidity-dependent changes in the focal adhesion dynamics can indeed be used to model the mechanism underpinning the persistence-stiffness correlation.

\begin{figure}[t!]
\includegraphics[width=1\columnwidth]{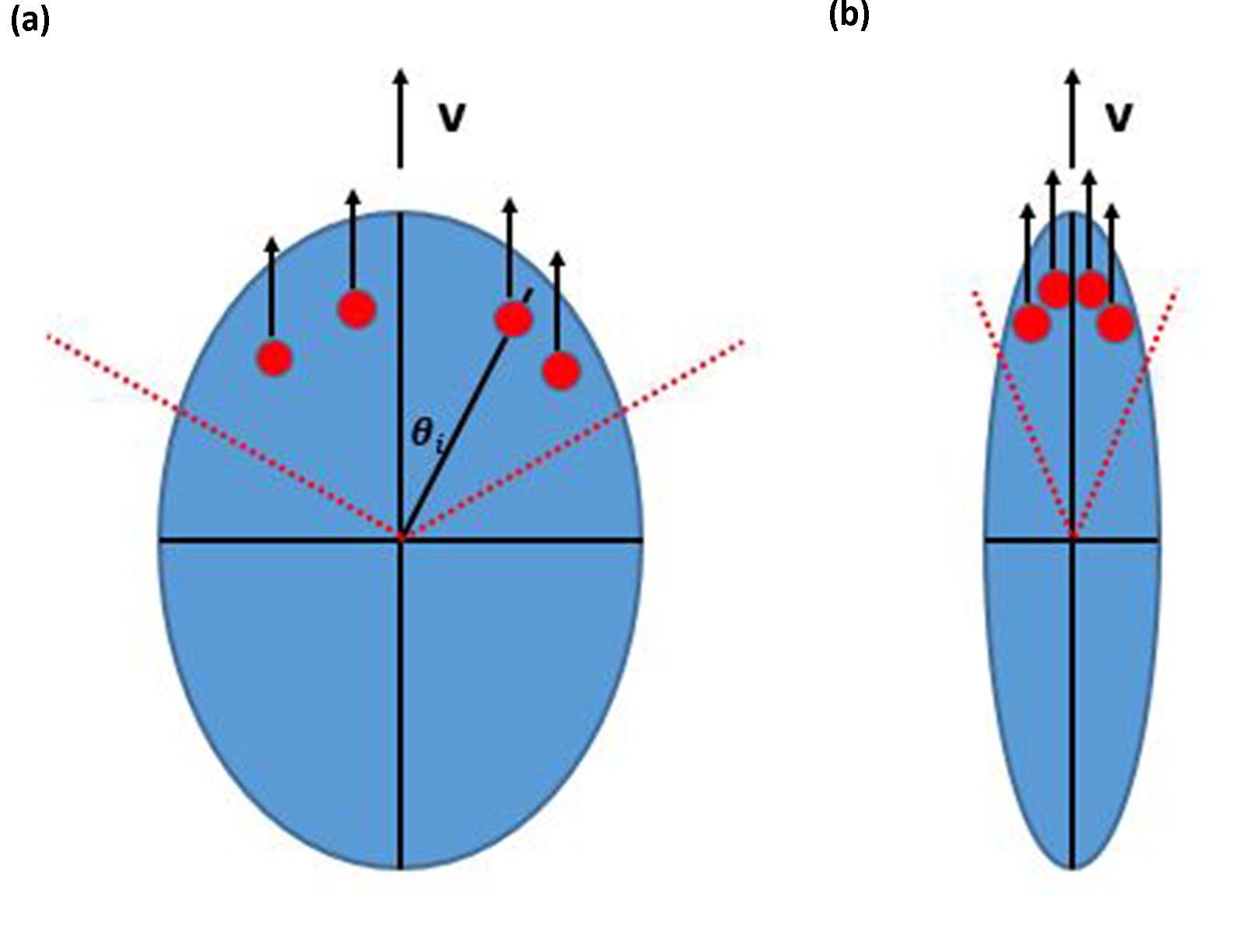}\caption{{\em A sketch of our model}. Red dots represent focal adhesions. In our simulation, focal adhesions are randomly generated within an angular range bounded by red lines in the figure at each step. The adhesions are all a constant distance from the cell center (a) For cells on a soft substrate, the distribution of FAs is relatively wide (b) Conversely,  for cells on a hard substrate, the distribution of FAs is relatively narrow. } \label{Fig1}
\end{figure}

Since focal adhesions are dynamically formed and destroyed, in our model at each time step the locations of all focal adhesions $\theta_i$ are reselected with no correlation to their previous value, hence  $<\Delta\Phi(t_i)\Delta\Phi(t_{i+1})>=0$.  Thus on a uniform substrate, approximating the distribution of $\Delta \Phi$ to be Gaussian with the calculated width reduces our model to a version of the worm-like chain, where the mean squared displacement is:
\be
 <x^2>=v^2\tau^2_p(\frac{t}{\tau_p}+e^{-\frac{t}{\tau_p}}-1)
\ee
Here the persistence time is defined as $\tau_p=-\frac{1}{v \ln <\cos\Delta\Phi>}$. Since $<\cos\Delta\Phi>=e^{-\frac{\sigma^2}{2}}$, where $\sigma^2=var(\Phi)$, $\tau_p=\frac{2}{v\sigma^2}$. We simulate cell trajectories on uniform substrates with different stiffness and verify the previous results for $\sigma^2$ (see Fig 2b). 
In Fig. 2c and 2d, we show trajectories of cells simulated on both uniform soft and hard substrates.  Consistent with the experimental observation \cite{RaabSwiftDingalEtAl2012}, cells crawl more efficiently on stiffer substrates.
\begin{figure}[t!]
\includegraphics[width=\columnwidth]{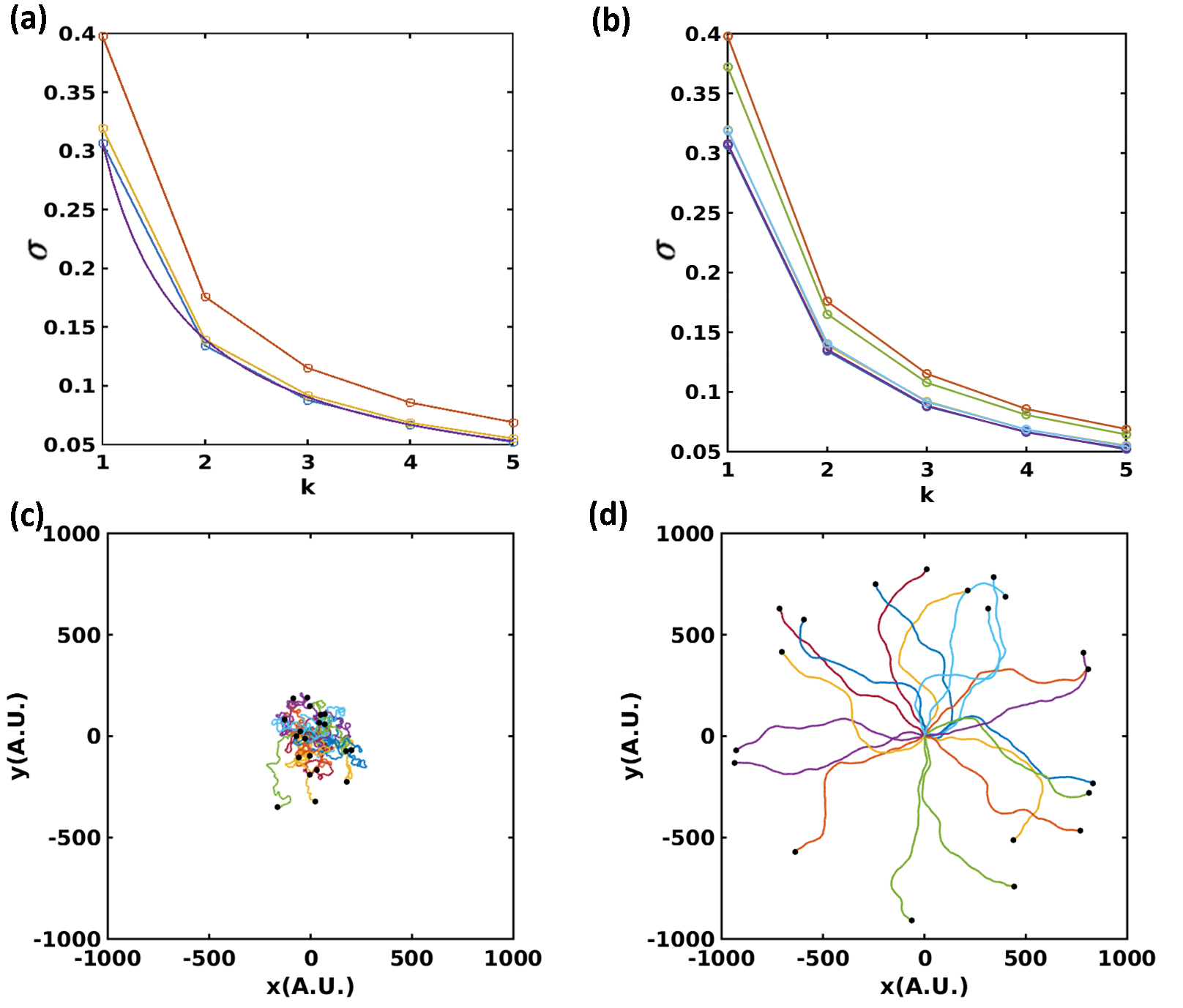}\caption{{\em Simulation result of uniform substrates and the variance of deflected angles.} 
(a) The variance of deflected angles calculated by Monte Carlo sampling. $\sigma^2$ vs stiffness $k$. $N=8, r_i=1(red)$, $N=12, r_i=1(blue)$, $N=12, r_i\in (0.5,1.5)$ (yellow) for all i and the fitted function $\sigma(k)$ with $\alpha=3.9$ and $\beta=-0.645 $(Eq 3)(purple). (b) Comparison between Monte Carlo sampling and direct simulation. For $N=12, r=1(red)$, two lines overlap together as the purple line. For $N=12, r_i\in (0.5,1.5)$, two lines overlap together as the blue line.$N=8, r_i=1(green) (direct simulation) $,$N=8, r_i=1(red) (Monte Carlo)$. For (c) and (d), we simulate 1000 time steps with $v=1 (A.U.)$  Initial position of 20 cells are (0,0) and the initial moving direction is randomly selected. The black dots are the final positions of each cell.  (c) On a soft substrate, k=1 and $\theta_i\in(-0.5\pi,0.5\pi)$ for all i. (d) On a stiff substrate with k=5, the angle range is $\theta_i\in(-0.1\pi,0.1\pi)$, for all i.} \label{Fig2}
\end{figure}

\begin{figure}
\includegraphics[width=1\columnwidth]{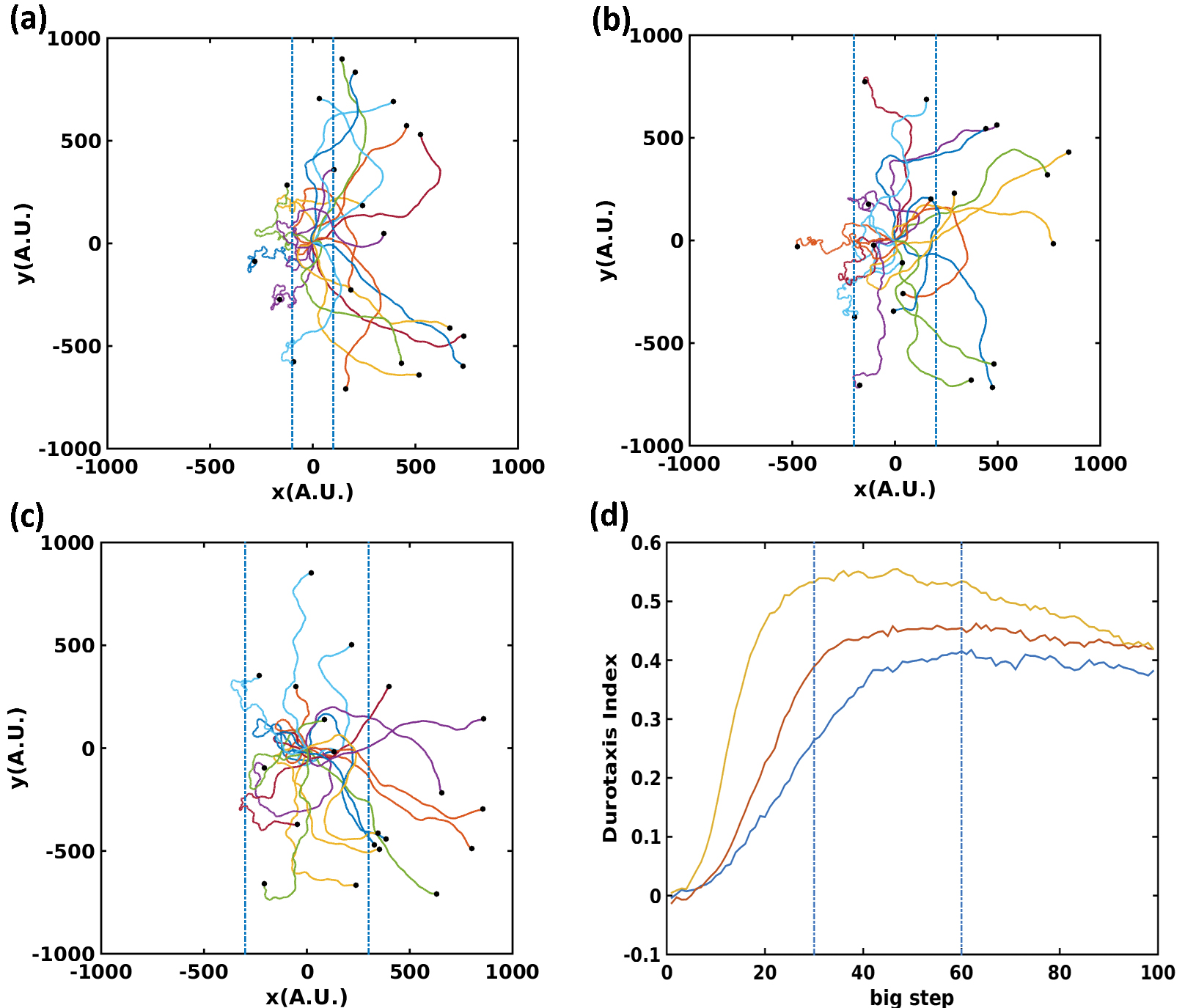}\caption{{\em Direct simulation on gradient matrix and DI. }(a)-(c) Soft substrate k=1 in the left region and hard substrate k=5 in the right region. The central region has a constant gradient stiffness and varying width L = (a) 100, (b) 200 and (c) 300 (A.U.). (d) Durotaxis Indices. Every 10 steps is counted as a "big step"} \label{Fig5}
\end{figure}

Next, we study the effect of stiffness gradients on cell motility. We set impose a constant stiffness gradient in the central region with constant low stiffness $k_{left}$ on the left side and high stiffness $k_{right}$ on the right side. We fix both $k_{left}$ and $k_{right}$ and vary the width of central region. 
\be
k(x)=\begin{cases}
k_{left}=1 & -1000<x<-L\\
k_{left} + \frac{(k_{right}-k_{left})}{2L}(X+L) & -L\leq X\leq L\\
k_{right}=5 & L<x<1000
\end{cases}
\ee
Initially all our cells are placed at the origin and given a random initial direction.
For small width, at a time when half of the cells go into the stiff region on the right, the other half are still hovering within the central gradient region (Fig. \ref{Fig5}a). As the width increases, fewer and fewer cells enter the soft region on the left (Fig. \ref{Fig5}b and c).  This is caused by the fact that larger width allows more moving steps inside the gradient region and cells have more time to adapt to the direction of stiffness gradient. We further characterize these results by the durotactic index (DI) (\cite{raab2012crawling}).  We calculate DI defined below every ten time steps in our simulation:
\be
Durotaxis\ Index(t_i)=\frac{N_{right}-N_{left}}{N_{right}+N_{left}}
\ee
where $N_{right}$ and $N_{left}$ are the number of cells instantaneously moving to the right and to the left respectively. This index ranges from $[-1,1]$. Larger Durotaxis Indices indicate more cells are moving towards the ascending gradient direction.

We find that the curve can be divided into three sections (Fig. \ref{Fig5}d). In the first section, nearly all cells are still in the gradient region. The index increases rapidly, which suggests that cells start being guided by the stiffness gradient. Consistent with experiment observation \cite{IsenbergDiMillaWalkerEtAl2009}, the magnitude of DI is highly correlated with the magnitude of the gradient.  In the second section, part of the cells are in the gradient region while the other have entered the uniform stiffness region. In the last section, the index starts decreasing because all cells move into the uniform rigid region and begin to execute random walks. The DI curve in the first section elucidates the role of gradient stiffness on cell motility.


To facilitate understanding of our simulation data, we now develop a Fokker-Plank equation for the probability distribution $P(x,y,\Phi ;t)$ governing a population of particles in our model. We will be specifically interested in cases with a stiffness gradient, which we choose to lie along the $x$ direction. We focus on the variation with $x$ and $\Phi$ and introduce  $p = \int dy P$ as a two-dimensional density. 
For any single cell, the next position $x(t+dt)$ depends on the current position and angle via $x(t)+v \cos(\Phi (t) )dt$. We can therefore represent a single step in our stochastic process via
\begin{equation}\label{eq4}
\begin{split}
p(x,\Phi ;t+dt) &=\intop_{0}^{2\pi}p(x-v \cos  (\Phi_{0}) dt,\Phi_{0};t) \\
 &f(x-v\cos (\Phi_{0}) dt,\Phi_{0}-\Phi) d \Phi_{0}
\end{split}
\end{equation}
where we will use the aforementioned Gaussian approximation
\begin{equation}
f(x,\Phi_{0}-\Phi)=a(x)e^{-\frac{(\Phi_{0}-\Phi)^{2}}{2\sigma(x)^{2} dt}}
\label{eq2}
\end{equation}
Here $a(x)=\sqrt{\frac{1}{2\pi\sigma(x)^{2}\mathrm{d}t}}$ is the normalization coefficient as long as the width is significantly smaller than $2\pi$. Note that now the variance depends on $x$ through an $x$ dependence in the stiffness $k$.

In standard manner we can assume small $dt$ and expand $p$ around the current values of its arguments. After some simplification, we obtain
\begin{equation}
\frac{\partial p}{\partial t}=-\frac{\partial p}{\partial x}v\ cos \Phi+\frac{\sigma(x)^{2}}{2}\frac{\partial^{2}p}{\partial\Phi^{2}}+v\ sin \Phi\frac{\partial}{\partial x}[\sigma(x)^{2}dt\frac{\partial p}{\partial\Phi}] \nonumber
\label{eq5}
\end{equation}
The horizontal location $x$ and moving direction $\Phi$ are directly coupled in the last term on the right side, which is of the order of $dt$. We have checked that this third term can be neglected for the discrete update steps in out simulation, Consequently, Eq(\ref{eq5}) can be simplified to
\begin{equation}
\frac{\partial p}{\partial t}=-\frac{\partial p}{\partial x}vcos(\Phi)+\frac{\sigma(x)^{2}}{2}\frac{\partial^{2}p}{\partial\Phi^{2}}
\label{eq6}
\end{equation}
This is of course the same as the Langevin equation for single cells:
\begin{equation}
\begin{split}
&\frac{dx}{dt}=v\ cos(\Phi)\\
&\frac{d\Phi}{dt}=\eta(t)
\label{eq7}
\end{split}
\end{equation}
where $<\eta(t)>=0$,$<\eta(t) \eta (t')>= \delta(t-t') \sigma(x)^2$. 
\begin{figure}[t]
\includegraphics[width=1\columnwidth]{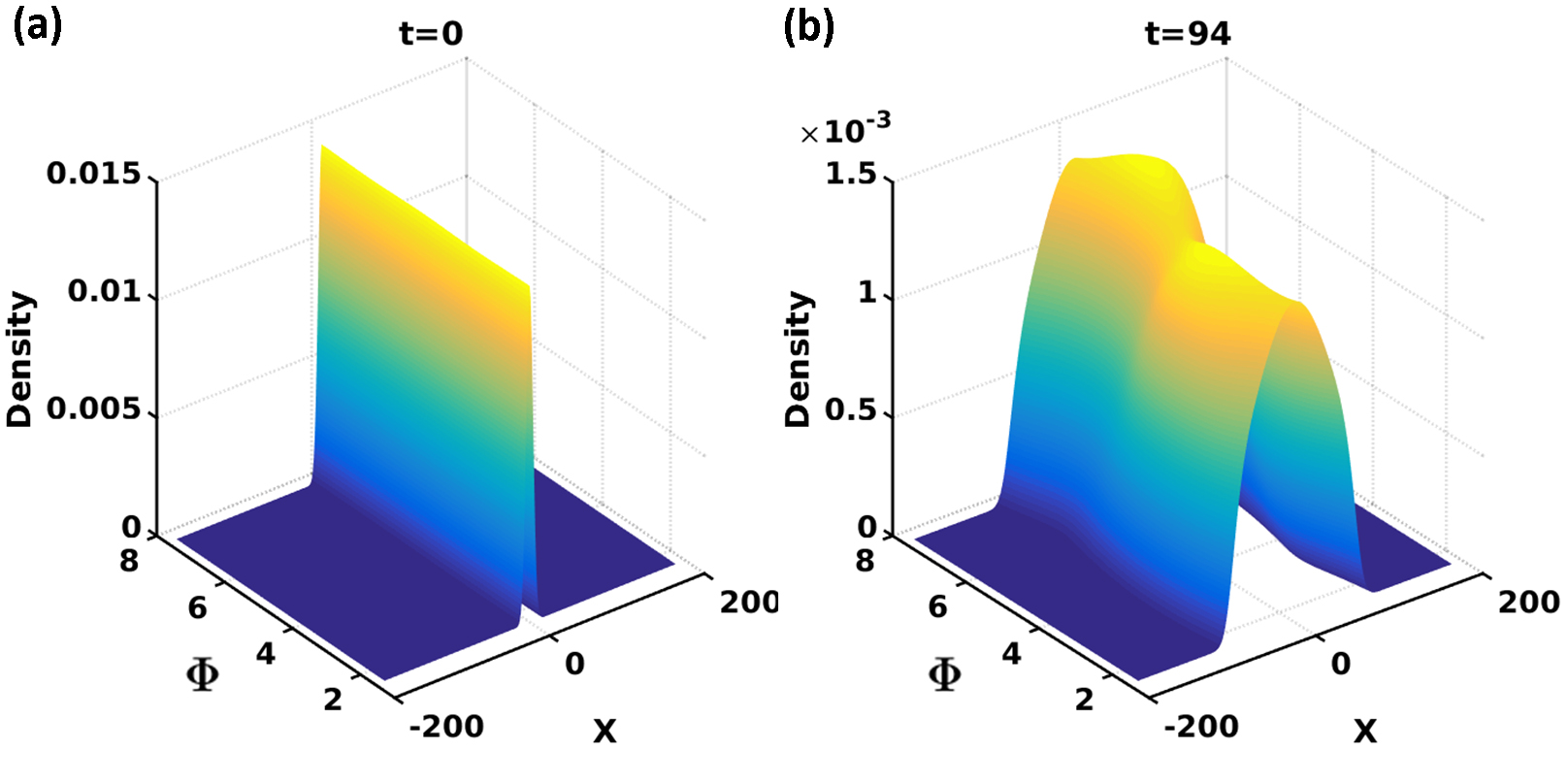}\caption{{\em PDE solution}. (a) Initial probability density function $p(x,\Phi; t=0)$ (b) Probability density function $p(x,\Phi; t=94)$ on a uniform substrate. Note that in this and the subsequent figure $\Phi$ runs from $\pi/2$ to $5\pi/2$. }
\end{figure}

\begin{figure}[t]
\includegraphics[width=1\columnwidth]{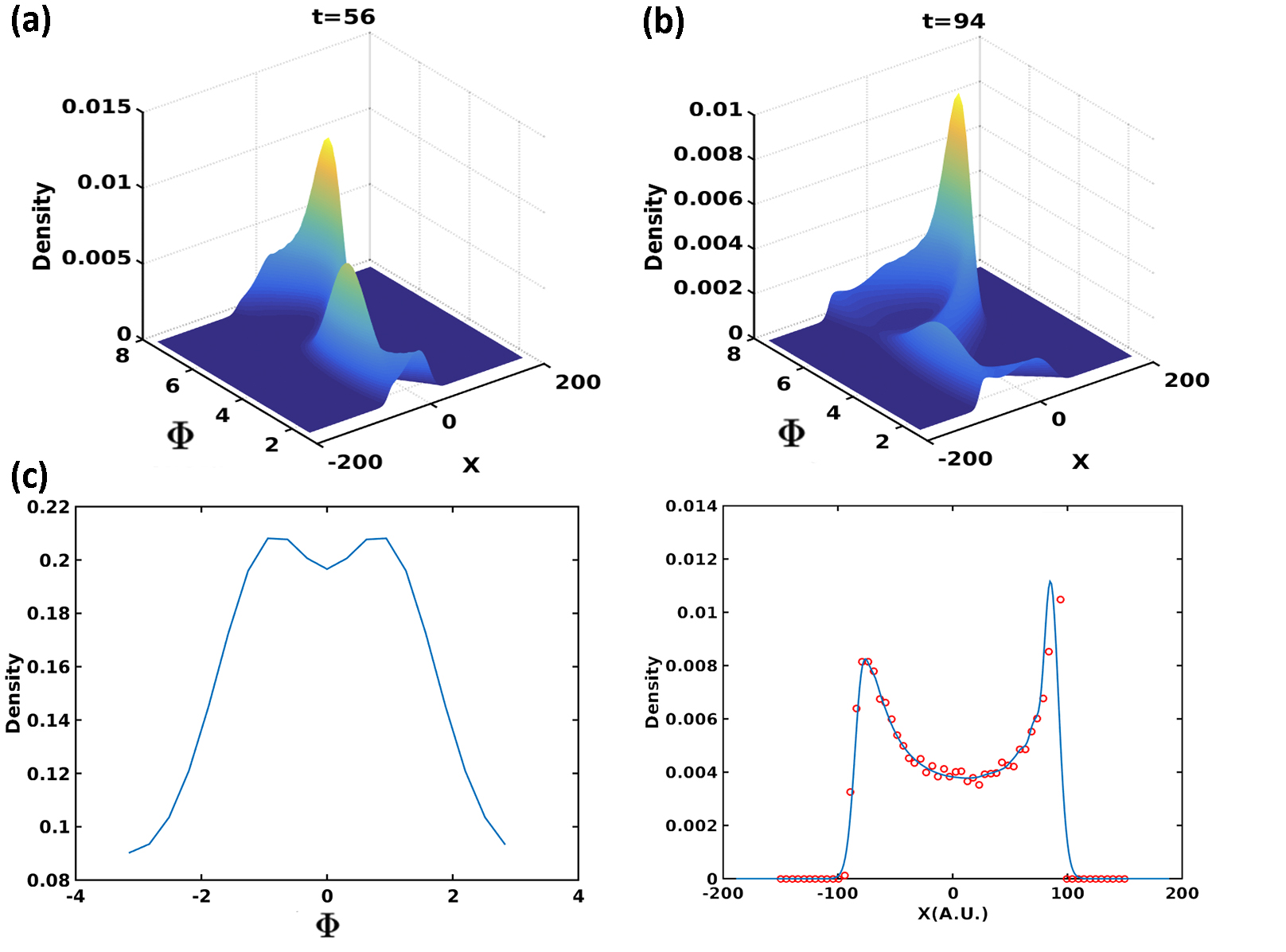}\caption{{\em PDE solution and comparison to simulation on a substrate with stiffness gradient}. Probability density distribution $p(x,\Phi; t)$ at (a) t=56 (b) t=94. (c) Distribution of the moving direction $\tilde{p}(\Phi$, t=94) (d) Comparison between direct simulations of 30000 cells and the numerical solution of the Fokker-Planck equation for $\hat{p}(x, t=94)$} \label{Fig4}
\end{figure}


We then solve Eq (9) for both the uniform and stiffness gradient substrate cases. Fig 4a shows the initial condition in the all cases discussed in the following; in particular we apply a narrow gaussian distribution to approximate the $\delta$ in $p(x, \Phi; t=0)=\frac{1}{2\pi}\delta(x), \Phi \in (0, 2\pi)$. The solution shows a peak in $x$ which varies from being at positive values (for $\Phi \simeq 0$) to negative ones (for $\Phi \simeq \pi$); the peak heights are independent of $\Phi$ as expected via rotational symmetry.

For the stiffness gradient case, the stiffness distribution is described by Eq (5) with $L=400$. The full distributions are shown at several times in Figs. 5a and 5b. Now there is a clear peak as a function of the direction; this can be seen most directly by plotting the integrated distribution $\tilde{p}(\Phi,t)=\intop_{-\infty}^{\infty}p(x,\Phi,t)dx$. Most cells adapt their moving directions from their initial directions $\Phi(t=0)$  to $\Phi$ near zero, exhibiting durotaxis (Fig 5c). We also define $\hat{p}(x,t)=\intop_{0}^{2\pi}p(x,\Phi,t)d\Phi$ and compare our PDE result for this quantity with direct simulations (Fig 5d). The very good agreement between PDE and direct simulation results validates the Fokker-Planck equation approach. Note that the above trends continue with the population continuing to break up into a peak at positive $x$ and a straggler peak  at negative $x$ corresponding to cells that have wandered out into the uniform less stiff side of the gradient profile.

In our calculations so far, we fixed the number of focal adhesions and varied their distributions. In fact, FA's are observed to be more stable on a stiffer substrate\cite{PelhamWang1997}, which suggests a larger number of effective FA's at any specific time. An alternative model would have a fixed FA distribution and vary the number of FA on substrates with different stiffness. In addition, it is possible that cells move with faster speed on stiffer substrates. Our model is capable of including such effects, all of which can contribute to durotaxis.

%


In this study, we discussed a possible underlying mechanism for durotaxis, namely a stiffness dependence of FA formation. It is known that FAs can dynamically sample rigidity to act as mechanosensors \cite{PlotnikovPasaperaSabassEtAl2012}, but it remains elusive how FAs formation can directly control cell motility. We show that stiffness dependent FA formation causes a positive correlation between persistence time and substrate stiffness, which leads to durotaxis \cite{NovikovaRaabDischerEtAl2015}. Also, we derive the corresponding 2D Fokker-Planck equation associated with our model and show that it gives consistent numerical agreement with our simulations. Our work can potentially help in predicting cell motility in more complex physiological environments such as those we arise during cancer metastasis.

Our model implicitly assumes that cells are incompetent in sensing rigidity gradients without moving around.  For chemotaxis, a close analog of durotaxis, a eukaryotic cell is capable of comparing chemical concentration between its two ends, even though a typical bacterium bacteria is not \cite{witzany2016biocommunication}. It is technically hard to test such an ability in durotaxis, mainly because the cytoskeleton is essential for both cell motility and mechanosensing.  Recently it has been shown that some cells can exhibit durotaxis as a cluster even if isolated constituent cells are ineffective \cite{SunyerConteEscribanoEtAl2016}; in this case motion is not necessary. 

{\em Acknowledgements}. This work was supported by the National Science Foundation Center for Theoretical Biological Physics (Grant NSF PHY-1427654). HL  was also supported by the CPRIT Scholar program of the State of Texas.

\bibliography{aa}

\begin{thebibliography}{10}

\bibitem{WangButlerIngberEtAl1993a}
Ning Wang, James~P Butler, Donald~E Ingber, et~al.
\newblock Mechanotransduction across the cell surface and through the
  cytoskeleton.
\newblock {\em Science}, 260(5111):1124--1127, 1993.

\bibitem{GuilakCohenEstesEtAl2009}
Farshid Guilak, Daniel~M Cohen, Bradley~T Estes, Jeffrey~M Gimble, Wolfgang
  Liedtke, and Christopher~S Chen.
\newblock Control of stem cell fate by physical interactions with the
  extracellular matrix.
\newblock {\em Cell stem cell}, 5(1):17--26, 2009.

\bibitem{ParkChuTsouEtAl2011}
Jennifer~S Park, Julia~S Chu, Anchi~D Tsou, Rokhaya Diop, Zhenyu Tang, Aijun
  Wang, and Song Li.
\newblock The effect of matrix stiffness on the differentiation of mesenchymal
  stem cells in response to tgf-$\beta$.
\newblock {\em Biomaterials}, 32(16):3921--3930, 2011.

\bibitem{EvansMinelliGentlemanEtAl2009}
Nicholas~D Evans, Caterina Minelli, Eileen Gentleman, Vanessa LaPointe,
  Sameer~N Patankar, Maria Kallivretaki, Xinyong Chen, Clive~J Roberts, and
  Molly~M Stevens.
\newblock Substrate stiffness affects early differentiation events in embryonic
  stem cells.
\newblock {\em Eur Cell Mater}, 18(1):e13, 2009.

\bibitem{TrappmannGautrotConnellyEtAl2012}
Britta Trappmann, Julien~E Gautrot, John~T Connelly, Daniel~GT Strange, Yuan
  Li, Michelle~L Oyen, Martien A~Cohen Stuart, Heike Boehm, Bojun Li, Viola
  Vogel, et~al.
\newblock Extracellular-matrix tethering regulates stem-cell fate.
\newblock {\em Nature materials}, 11(7):642--649, 2012.

\bibitem{PelhamWang1997}
Robert~J Pelham and Yu-li Wang.
\newblock Cell locomotion and focal adhesions are regulated by substrate
  flexibility.
\newblock {\em Proceedings of the National Academy of Sciences},
  94(25):13661--13665, 1997.

\bibitem{YeungGeorgesFlanaganEtAl2005}
Tony Yeung, Penelope~C Georges, Lisa~A Flanagan, Beatrice Marg, Miguelina
  Ortiz, Makoto Funaki, Nastaran Zahir, Wenyu Ming, Valerie Weaver, and Paul~A
  Janmey.
\newblock Effects of substrate stiffness on cell morphology, cytoskeletal
  structure, and adhesion.
\newblock {\em Cell motility and the cytoskeleton}, 60(1):24--34, 2005.

\bibitem{LoWangDemboEtAl2000}
Chun-Min Lo, Hong-Bei Wang, Micah Dembo, and Yu-li Wang.
\newblock Cell movement is guided by the rigidity of the substrate.
\newblock {\em Biophysical journal}, 79(1):144--152, 2000.

\bibitem{Basan2013}
Markus Basan, Jens Elgeti, Edouard Hannezo, Wouter-Jan Rappel, and Herbert
  Levine.
\newblock Alignment of cellular motility forces with tissue flow as a mechanism
  for efficient wound healing.
\newblock {\em Proceedings of the National Academy of Sciences},
  110(7):2452--2459, 2013.

\bibitem{Losick2013}
Vicki~P Losick, Donald~T Fox, and Allan~C Spradling.
\newblock Polyploidization and cell fusion contribute to wound healing in the
  adult drosophila epithelium.
\newblock {\em Current Biology}, 23(22):2224--2232, 2013.

\bibitem{Huang2005}
Sui Huang and Donald~E Ingber.
\newblock Cell tension, matrix mechanics, and cancer development.
\newblock {\em Cancer cell}, 8(3):175--176, 2005.

\bibitem{Venkatesh2008}
Sudhakar~K Venkatesh, Meng Yin, James~F Glockner, Naoki Takahashi, Philip~A
  Araoz, Jayant~A Talwalkar, and Richard~L Ehman.
\newblock Mr elastography of liver tumors: preliminary results.
\newblock {\em American Journal of Roentgenology}, 190(6):1534--1540, 2008.

\bibitem{SambethBaumgaertner2001}
R~Sambeth and A~Baumgaertner.
\newblock Autocatalytic polymerization generates persistent random walk of
  crawling cells.
\newblock {\em Physical review letters}, 86(22):5196, 2001.

\bibitem{HuangBrangwynneParkerEtAl2005}
S~Huang, CP~Brangwynne, KK~Parker, and DE~Ingber.
\newblock Symmetry-breaking in mammalian cell cohort migration during tissue
  pattern formation: Role of random-walk persistence.
\newblock {\em Cell motility and the cytoskeleton}, 61(4):201--213, 2005.

\bibitem{SadjadiShaebaniRiegerEtAl2015a}
Zeinab Sadjadi, M~Reza Shaebani, Heiko Rieger, and Ludger Santen.
\newblock Persistent-random-walk approach to anomalous transport of
  self-propelled particles.
\newblock {\em Physical Review E}, 91(6):062715, 2015.

\bibitem{Reynolds2010}
AM~Reynolds.
\newblock Can spontaneous cell movements be modelled as l{\'e}vy walks?
\newblock {\em Physica A: Statistical Mechanics and its Applications},
  389(2):273--277, 2010.

\bibitem{BuldyrevGoldbergerHavlinEtAl1993}
Sergey~V Buldyrev, Ary~L Goldberger, Shlomo Havlin, Chung-Kang Peng, Michael
  Simons, and H~Eugene Stanley.
\newblock Generalized l{\'e}vy-walk model for dna nucleotide sequences.
\newblock {\em Physical Review E}, 47(6):4514, 1993.

\bibitem{NovikovaRaabDischerEtAl2015}
Elizaveta~A Novikova, Matthew Raab, Dennis~E Discher, and Cornelis Storm.
\newblock Persistence-driven durotaxis: Generic, directed motility in rigidity
  gradients.
\newblock {\em Physical Review Letters}, 118(7):078103, 2017.

\bibitem{FacklerGrosse2008}
Oliver~T Fackler and Robert Grosse.
\newblock Cell motility through plasma membrane blebbing.
\newblock {\em The Journal of cell biology}, 181(6):879--884, 2008.

\bibitem{Keren2011}
Kinneret Keren.
\newblock Cell motility: the integrating role of the plasma membrane.
\newblock {\em European Biophysics Journal}, 40(9):1013--1027, 2011.

\bibitem{GalkinOrlovaEgelman2012}
Vitold~E Galkin, Albina Orlova, and Edward~H Egelman.
\newblock Actin filaments as tension sensors.
\newblock {\em Current Biology}, 22(3):R96--R101, 2012.

\bibitem{MattilaLappalainen2008}
Pieta~K Mattila and Pekka Lappalainen.
\newblock Filopodia: molecular architecture and cellular functions.
\newblock {\em Nature reviews Molecular cell biology}, 9(6):446--454, 2008.

\bibitem{KimWirtz2013}
Dong-Hwee Kim and Denis Wirtz.
\newblock Focal adhesion size uniquely predicts cell migration.
\newblock {\em The FASEB Journal}, 27(4):1351--1361, 2013.

\bibitem{PlotnikovPasaperaSabassEtAl2012}
Sergey~V Plotnikov, Ana~M Pasapera, Benedikt Sabass, and Clare~M Waterman.
\newblock Force fluctuations within focal adhesions mediate ecm-rigidity
  sensing to guide directed cell migration.
\newblock {\em Cell}, 151(7):1513--1527, 2012.

\bibitem{trichet2012evidence}
L{\'e}a Trichet, Jimmy Le~Digabel, Rhoda~J Hawkins, Sri Ram~Krishna Vedula,
  Mukund Gupta, Claire Ribrault, Pascal Hersen, Rapha{\"e}l Voituriez, and
  Beno{\^\i}t Ladoux.
\newblock Evidence of a large-scale mechanosensing mechanism for cellular
  adaptation to substrate stiffness.
\newblock {\em Proceedings of the National Academy of Sciences},
  109(18):6933--6938, 2012.

\bibitem{IsenbergDiMillaWalkerEtAl2009}
Brett~C Isenberg, Paul~A DiMilla, Matthew Walker, Sooyoung Kim, and Joyce~Y
  Wong.
\newblock Vascular smooth muscle cell durotaxis depends on substrate stiffness
  gradient strength.
\newblock {\em Biophysical journal}, 97(5):1313--1322, 2009.

\bibitem{RaabSwiftDingalEtAl2012}
Matthew Raab, Joe Swift, PC~Dave~P Dingal, Palak Shah, Jae-Won Shin, and
  Dennis~E Discher.
\newblock Crawling from soft to stiff matrix polarizes the cytoskeleton and
  phosphoregulates myosin-ii heavy chain.
\newblock {\em The Journal of cell biology}, 199(4):669--683, 2012.

\bibitem{raab2012crawling}
Matthew Raab, Joe Swift, PC~Dave~P Dingal, Palak Shah, Jae-Won Shin, and
  Dennis~E Discher.
\newblock Crawling from soft to stiff matrix polarizes the cytoskeleton and
  phosphoregulates myosin-ii heavy chain.
\newblock {\em The Journal of cell biology}, 199(4):669--683, 2012.

\bibitem{witzany2016biocommunication}
Guenther Witzany and Mariusz Nowacki.
\newblock {\em Biocommunication of ciliates}.
\newblock Springer, 2016.

\bibitem{SunyerConteEscribanoEtAl2016}
Raimon Sunyer, Vito Conte, Jorge Escribano, Alberto Elosegui-Artola, Anna
  Labernadie, L{\'e}o Valon, Daniel Navajas, Jos{\'e}~Manuel Garc{\'\i}a-Aznar,
  Jos{\'e}~J Mu{\~n}oz, Pere Roca-Cusachs, et~al.
\newblock Collective cell durotaxis emerges from long-range intercellular force
  transmission.
\newblock {\em Science}, 353(6304):1157--1161, 2016.

\end{thebibliography}

\end{document}